\documentclass[reprint, aps, prl, amsmath, amssymb, superscriptaddress]{revtex4-2}

\usepackage{graphicx}
\usepackage{physics}
\usepackage{bm}
\usepackage{multirow}
\usepackage{url}

\usepackage{xcolor}

\newcommand{\diagarrow}{\mathrel{\rotatebox[origin=c]{45}{$\leftrightarrow$}}}

\begin{document}


\title[Extracting Anyon Statistics from Neural Network Fractional Quantum Hall States]{Extracting Anyon Statistics from Neural Network Fractional Quantum Hall States} 



\author{Andres Perez Fadon}
\email{andres.perez-fadon19@imperial.ac.uk}
\affiliation{Department of Physics, Imperial College London, London SW7 2AZ, United Kingdom}

\author{David Pfau}
\author{James S. Spencer}
\affiliation{DeepMind, London N1C 4DJ, United Kingdom}

\author{Wan Tong Lou}
\affiliation{Department of Physics, Imperial College London, London SW7 2AZ, United Kingdom}

\author{Titus Neupert}
\affiliation{Department of Physics, University of Z\"urich, Winterthurerstrasse 190, 8057 Z\"urich, Switzerland}

\author{W. M. C. Foulkes}
\affiliation{Department of Physics, Imperial College London, London SW7 2AZ, United Kingdom}


\date{\today}

\begin{abstract}
Fractional quantum Hall states host emergent anyons with exotic exchange statistics, but obtaining direct access to their topological properties in real systems remains a challenge. 
Neural-network wavefunctions provide a flexible computational approach, as they can represent highly correlated states without requiring a tailored basis. 
Here we use the neural-network variational Monte Carlo method to study the fractional quantum Hall effect on the torus and find the three degenerate ground states at filling factor $\nu=1/3$. 
From these, we extract the modular $S$ matrix via entanglement interferometry, a technique previously only applied to lattice models.
The resulting $S$ matrix encodes the quantum dimensions, fusion rules, and exchange statistics of the emergent anyons, providing a direct numerical demonstration of the topological order. 
The calculated anyon properties match the well-known theoretical and experimental results.
Our work establishes neural-network wavefunctions as a powerful new tool for investigating anyonic properties.
\end{abstract}


\maketitle 



The fractional quantum Hall effect (FQHE)~\cite{tsui1982two, laughlin1983anomalous, haldane1983fractional, arovas1984fractional, haldane1985finite, jain1989composite, moore1991nonabelions, rezayi2000incompressible, jain2007composite} arises when a strong magnetic field is applied to a two-dimensional layer of interacting electrons with a fractional Landau level filling factor $\nu$. 
While the integer quantum Hall effect can be understood in terms of non-interacting electrons, the FQHE is intrinsically interaction driven and hosts anyonic excitations with fractional charge~\cite{laughlin1983anomalous, de1998direct, saminadayar1997observation} and exchange statistics~\cite{arovas1984fractional, wilczek1982magnetic}, which may be Abelian or non-Abelian~\cite{moore1991nonabelions, greiter1992paired, rezayi2000incompressible, ma2022fractional}. 
The latter are of particular interest because of their potential use as decoherence-resistant qubits in topological quantum computing~\cite{nayak2008non, kitaev2003fault, georgiev2017topological}.

\begin{figure*}[!t] 
    \centering
    \includegraphics[width=\textwidth]{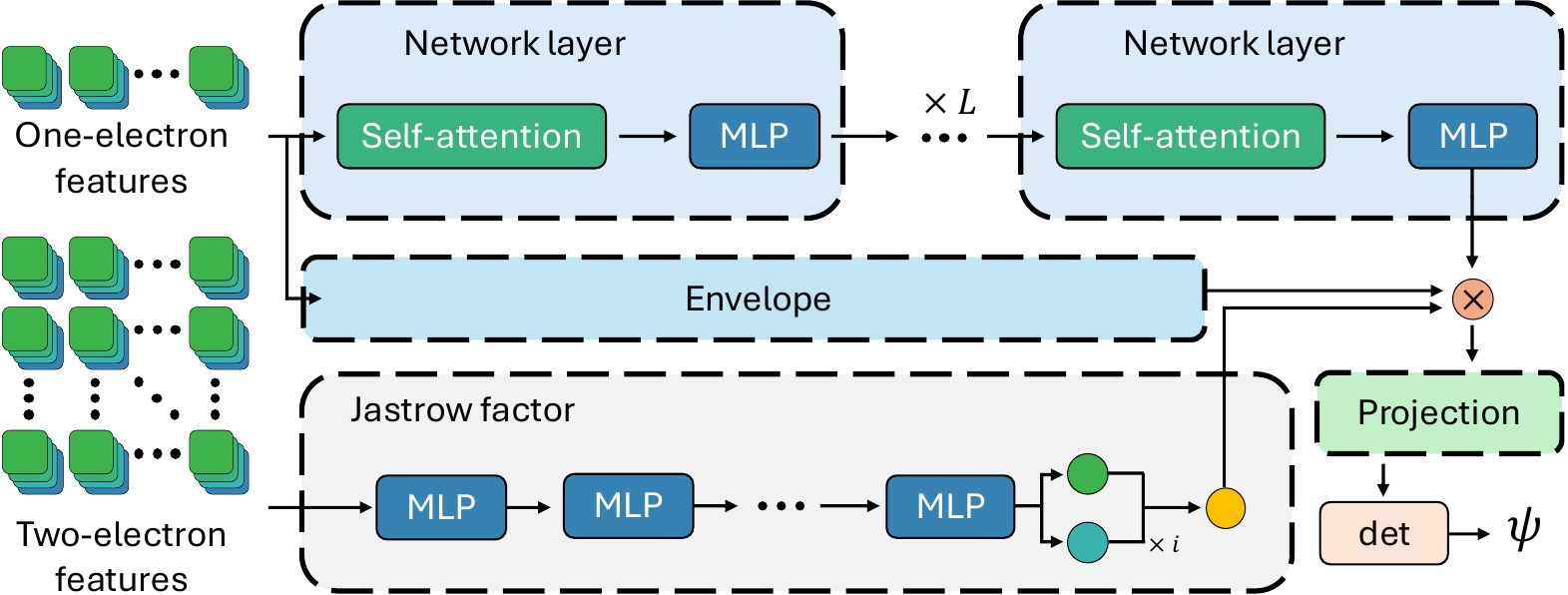}
    \caption{Psiformer architecture with a custom complex-valued multi-layer-perceptron-based Jastrow factor.
      The one-electron features depend on periodic functions of the position of each electron~\cite{cassella2023discovering}.
    The Jastrow factor is evaluated independently for each pair of electrons.
    The values from each pair are then multiplied together.
    The envelope is a sum of quasiperiodic functions with learnable coefficients.}
    \label{fig:network}
\end{figure*}

Computational studies of the FQHE have used exact diagonalization, density-matrix renormalization group (DMRG) methods and variational trial wavefunctions. 
Most calculations are restricted to the lowest Landau level, either on the surface of a sphere~\cite{morf1998transition, wan2008fractional, feiguin2008density, kovrizhin2010density, zhao2011fractional, arciniaga2016landau, zuo2020interplay, qian2025describing} or in the plane with a disk of positive charge binding the electrons~\cite{laughlin1983anomalous, morf1986monte, wan2008fractional, hu2012comparison, teng2025solving}.
However, the assumption of vanishing Landau level mixing is not always valid, as fractional quantum Hall states have been seen for dimensionless mixing parameters up to $\kappa=2.8$~\cite{samkharadze2017observation}.
In this limit, the assumption of vanishing Landau level mixing does not even approximately hold.

In recent years, neural wavefunctions~\cite{carleo2017solving, choo2020fermionic, hermann2020deep, pfau2020ab, spencer2020better, gao2021ab, li2022ab, li2022fermionic, pescia2022neural, qian2022interatomic, scherbela2022solving, von2022self, cassella2023discovering, entwistle2023electronic, gao2023generalizing, hermann2023ab, cassella2024neural, li2024electric, lou2024neural, pescia2024message, qian2024force, scherbela2024towards, perez2025interaction} have proven able to provide accurate descriptions of a wide variety of interacting electron systems.
Many of the aforementioned examples are strongly correlated and difficult to study with standard electronic structure methods because one lacks a many-body basis set that accurately captures the underlying physical phenomena, such as virtual positronium formation in positronic chemistry \cite{cassella2024neural}.
Neural networks, by contrast, are such expressive function approximators that, with only minor modifications, the same network architecture is capable of representing a wide range of correlated states accurately.
Here we show that neural wavefunctions are able to describe the topological order of quantum Hall states and can be used to study the properties of the emergent anyons.

The most interesting property of the FQHE ground state is that it has fractional excitations. The topological properties of these excitations are hard to investigate using direct computational methods, but
several indirect approaches have been used to verify the topological order of fractional quantum Hall states obtained numerically.
The first is to calculate the overlaps of the wavefunction with known topological states~\cite{morf1998transition, peterson2015abelian, pakrouski2015phase, jeong2015bilayer, rezayi2017landau}, on the assumption that the topological order is shared with the state that has the highest overlap.
The second is to demonstrate an adiabatic connection.
One can check numerically that the many-body gap remains open as, for instance, the Coulomb Hamiltonian $H_{0}$ is smoothly deformed to a Hamiltonian $H_{1}$ with a ground state of known topological order~\cite{storni2010fractional, hutzel2019particle}.
If this is true, the topological order of the ground state of $H_{0}$ is the same as that of the ground state of $H_{1}$.
Entanglement entropy and in particular the entanglement spectrum contain fingerprints of the topological order~\cite{li2008entanglement, mong2017fibonacci}. Lastly, some studies have calculated the braid phase by explicitly braiding the anyons~\cite{prodan2009mapping, storni2011localized}.

This Letter reports the results of a neural-network-based variational Monte Carlo (NNVMC) study of the fractional quantum Hall effect at $1/3$ filling on the torus.
We calculate the three degenerate ground states and use them to directly obtain the modular $S$ matrix~\cite{zhang2012quasiparticle, cincio2013characterizing, zhu2013minimal, zhu2014identifying, zhang2015general, li2022detecting}, which encodes the topological properties of the emergent anyons~\cite{bakalov2001lectures, kassel2012quantum, turaev2010quantum, simon2023topological}.
From the modular $S$ matrix, we calculate the quantum dimensions, fusion rules, and braid phase of the anyons.
This is accomplished using entanglement interferometry, a method previously used for lattice systems~\cite{zhang2012quasiparticle, zhang2015general}.

Two other recent papers have used neural wavefunctions to simulate the FQHE in the continuum~\cite{qian2025describing, teng2025solving}, but did not extract properties of the topological order.
Lattice analogues of the FQHE, so-called fractional Chern insulators, have also been studied with neural network quantum states~\cite{li2025deep}, but again the topological properties of the resulting wavefunctions were not considered.


The neural wavefunction we employ is based on the Psiformer architecture~\cite{von2022self} as implemented in the FermiNet package~\cite{ferminet_github}, with most hyperparameters set to the default values.
(For a complete list, see the supplementary material.)
The Coulomb interactions were evaluated using 2D Ewald summations.
We take inspiration from the Laughlin wavefunction, which can be written as 
\begin{align}
    \psi_{\text{Laughlin}}(z_{1}, \dots, z_{N}) &= \widetilde{J} \psi_{\text{I}}(z_{1}, \dots, z_{N}),\\
    \widetilde{J} &= \prod_{i<j}^{N}(z_{i} - z_{j})^{2p},
\end{align}
where $N$ is the number of electrons, $2p = m - 1$ for filling factor $\nu = 1/m$ with $m$ odd, and $z_{j} = x_{j} + iy_{j}$.
The first term, $\widetilde{J}$, is often is often called a Jastrow factor, although it is complex valued and does not have the same exponential form as conventional Jastrow factors.
The second term is the solution of the non-interacting integer quantum Hall problem.
Analogously, we write our variational ansatz as 
\begin{align}
    \psi(\bm{r}_{1}, \dots, \bm{r}_{N}) &= J\sum_{d = 1}^{N_{\text{det}}}\det_{ij}[\phi^{d}_{i}(\bm{r}_{j}, \{\bm{r}_{/j}\})],
\end{align}
where $J$ is a complex-valued Jastrow factor and the second term is the output of the Psiformer~\cite{von2022self,ferminet_github} (with the default Jastrow factor removed).
The structure of the wavefunction is shown schematically in Fig.~\ref{fig:network}.

The orbitals $\phi_{i}^{d}(\bm{r}_{j}, \{\bm{r}_{/j}\})$ are expressed as
\begin{align}
    \phi_{i}^{d}(\bm{r}_{j}, \{\bm{r}_{/j}\}) &= \varphi_{i}^{d}(\bm{r}_{j}, \{\bm{r}_{/j}\})\chi_{i}^{d}(\bm{r}_{j}),
\end{align}
where $\varphi_{i}^{d}$ is a periodic function of all inputs obtained from the last layer of the network and $\chi_{i}^{d}$ is an envelope used to enforce the boundary conditions.
We work in the Landau gauge, $\bm{A} = (yB, 0)^{T}$, with the magnetic field pointing into the $xy$ plane. 
In the torus, this requires the use of quasi-periodic boundary conditions,
\begin{align}
    \psi(x + L_{x}, y) &= \psi(x, y),\\
    \psi(x, y + L_{y}) &= e^{-i2\pi N_{\phi}x/L_{x}}\psi(x, y),
\end{align}
where $L_x$ and $L_y$ are the lengths of the sides of the toroidal simulation cell and $N_{\phi}=\frac{1}{\nu}N\in \mathbb{Z}$ is the number of flux quanta piercing the simulation cell.
We write the envelope as a sum of quasi-periodic functions,
\begin{align}
    \chi^{d}_{i}(x, y) &= \sum_{n=0}^{N_{\text{LL}} - 1}\sum_{s=0}^{N_{\phi}-1}\sigma_{ns}^{di}e_{ns}(x, y),\\
    e_{ns}(x, y) &= \sum_{k}f_{ns}\left(\frac{y}{L_{y}} + k\right)e^{i2\pi k N_{\phi}x/L_{x}},\\
    f_{ns}(a)&=e^{-\pi N_{\phi}(a + s/N_{\phi})^{2}}H_{n}\left(a + s/N_{\phi}\right),
\end{align}
where $\sigma_{ns}^{di}$ are learnable parameters, $H_{n}$ is the $n$th Hermite polynomial, and $N_{\text{LL}}$ is a cut-off of Landau levels included. 
The form of the envelope is based on the eigenfunctions of a single particle moving on the torus subject to a magnetic field~\cite{fremling2019modular}.
The complex-valued Jastrow factor $J$ takes the form
\begin{align}
    J(\bm{r}_{1}, \dots, \bm{r}_{N}) &= \prod_{ij=1}^{N}\text{MLP}(\bm{s}_{i}, \bm{s}_{j}, \bm{s}_{ij}, s_{ij}),
\end{align}
where MLP stands for multi-layer perceptron, and $\bm{s}_{i}$ is a periodic function of $\bm{r}_{i}$~\cite{cassella2023discovering},
\begin{align}
    \bm{s}_i &= \left(\cos\left(\frac{2\pi}{L}\bm{r}_i\right), \sin\left(\frac{2\pi}{L}\bm{r}_{i}\right)\right),\\
    \bm{s}_{ij} &= \left(\cos\left(\frac{2\pi}{L}(\bm{r}_{i}-\bm{r}_{j})\right), \sin\left(\frac{2\pi}{L}(\bm{r}_{i}-\bm{r}_{j})\right)\right),\\
    s_{ij} &=\nonumber\\
    \sum_{ab}&\left[1-\cos\left(\frac{2\pi}{L_{a}}(r_{ij})_{a}\right)\right]M_{ab}\left[1-\cos\left(\frac{2\pi}{L_{b}}(r_{ij})_{b}\right)\right]\nonumber\\
    +&\sin\left(\frac{2\pi}{L_{a}}(r_{ij})_{a}\right)M_{ab}\sin\left(\frac{2\pi}{L_{b}}(r_{ij})_{b}\right),
\end{align}
where $a,b\in \{x, y\}$, and $M_{ab}=\bm{a}_{a} \cdot \bm{a}_{b}$, with $\bm{a}_{j}$ being the $j$th lattice vector of the torus.
The fact that the Jastrow factor returns a complex output is essential, as it enables the network to represent the complex phase structure of the quantum Hall state.

As described above, the method used to calculate the quasiparticle properties requires the computation of multiple degenerate ground states.
To maintain orthogonality, we use a penalty function method originally introduced for computing excited states~\cite{wheeler2024ensemble}.
The loss function is given by
\begin{align}
    \mathcal{L}[\{\psi_{i}\}] &= \sum_{i}w_{i}E[\psi_{i}] + \lambda \sum_{i<j}|\mathcal{O}_{ij}|^{2},
\end{align}
where $E[\psi]$ is the energy expectation value of $\psi$, $w_{i} \propto 1/i$, $\lambda$ was chosen to be unity, and $\mathcal{O}_{ij}$ is the overlap between $\psi_{i}$ and $\psi_{j}$.
As the method used to calculate the anyon properties requires access to the orthogonal ground states, the penalty method is more convenient than natural excited states~\cite{pfau2024accurate}, which would require further orthogonalization of the states.
The parameters were optimized with the KFAC algorithm~\cite{martens2015optimizing}.

\begin{table*}[!t]
\centering
\begin{tabular}{|c c c|c c c c|}
  \hline
  Ansatz & $N$ & $B$ & $E_{0}$ & $E_{1}$ & $E_{2}$ & $E_{4}$\\
  \hline
  FN+J & 4 & 10 & \bf{14.741808(5)} & \bf{14.741909(6)} & \bf{14.742021(7)} & \bf{14.879138(9)}\\
  PS+J & 4 & 10 & 14.741847(6) & 14.741960(7) & 14.742222(7) & 14.87914(1)\\
  Laughlin & 4 & 10 & 14.76114(2) & 14.76114(2) & 14.76114(2) & -\\
  \hline
  FN+J & 5 & 5 & 7.88224(1) & 7.88234(1) & 7.88243(1) & -\\
  PS+J & 5 & 5 & \bf{7.881565(9)} & \bf{7.881872(9)} & \bf{7.88207(1)} & -\\
  Laughlin & 5 & 5 & 7.88923(2) & 7.88923(2) & 7.88923(2) & -\\
  \hline
  FN+J & 6 & 5 & 9.47078(2) & 9.47096(3) & 9.47210(3) & -\\
  PS+J & 6 & 5 & \bf{9.46923(2)} & \bf{9.46935(2)} & \bf{9.46988(2)} & -\\
  Laughlin & 6 & 5 & 9.47803(2) & 9.47803(2) & 9.47803(2) & -\\
  \hline
  FN+J & 8 & 5 & 12.64981(5) & 12.65420(5) & 12.66193(7) & -\\
  PS+J & 8 & 5 & \bf{12.63763(3)} & \bf{12.63787(3)} & \bf{12.63806(3)} & -\\
  Laughlin & 8 & 5 & 12.64935(2) & 12.64935(2) & 12.64935(2) & -\\
  \hline
\end{tabular}
\caption{Variational energies (in Hartrees) obtained for the three lowest energy eigenstates (lowest energies in bold). 
We use the shorthand FN+J to denote the combination of a FermiNet wave function with a complex MLP Jastrow factor, and PS+J to denote the combination of a PsiFormer with a complex MLP Jastrow factor. 
$N$ is the number of electrons and $B$ is the magnetic field in atomic units. 
For the 4-electron case, we also calculated the energy of the 4$^{\text{th}}$ state, verifying that the ground state is three-fold degenerate. \label{table:energies}
}
\end{table*}

The energies obtained from our calculations are reported in Table \ref{table:energies}.
For the larger systems, the Psiformer~\cite{von2022self} consistently obtains a lower energy than the earlier FermiNet architecture \cite{pfau2020ab}.
Using the variational principle to optimize neural network representations of FQHE wave functions proceeds smoothly but takes a surprisingly large amount of computer time, see the supplementary material.
This prevented us from studying systems with more than 8 electrons.
Previous neural wavefunction studies of the FQHE~\cite{qian2025describing, teng2025solving} were similarly limited.
Further, these studies computed a single ground state in the sphere and disk geometry, respectively.
The calculation of the entire degenerate space in the torus is a significant step up in complexity.


We are interested in the modular $S$ matrix, which contains topological information about the anyon excitations of the Hamiltonian.
For example, the ground state degeneracy of a system with $N_{a}$ anyons of type $a$ approaches $d_{a}^{N_{a}}$ for large $N_{a}$, where $d_{a}$ is the quantum dimension of anyon type $a$.
The total quantum dimension $D$ is then defined by $D^{2}=\sum_{a}d_{a}^{2}$, where the sum runs over all allowed anyon types in the theory.
If one calculates the modular $S$ matrix, the quantum dimensions can be read from $S_{1a} = d_{a}/D$.
Further, as there is always a ``vacuum'' particle with quantum dimension 1 (call it $a=e$), the total quantum dimension is $D = 1 / S_{ee}$.
The modular $S$ matrix can also be used to work out the fusion rules of the anyons via the Verlinde formula (see Supplementary Material).
Further, if $a$ and $b$ are Abelian anyons, then $\arg(S_{ab})$ is the argument $\theta_{ab}$ of the complex phase obtained when braiding anyon $a$ around anyon $b$~\cite{delmastro2021symmetries}.

\begin{figure}
    \centering
    \includegraphics[scale = 0.25]{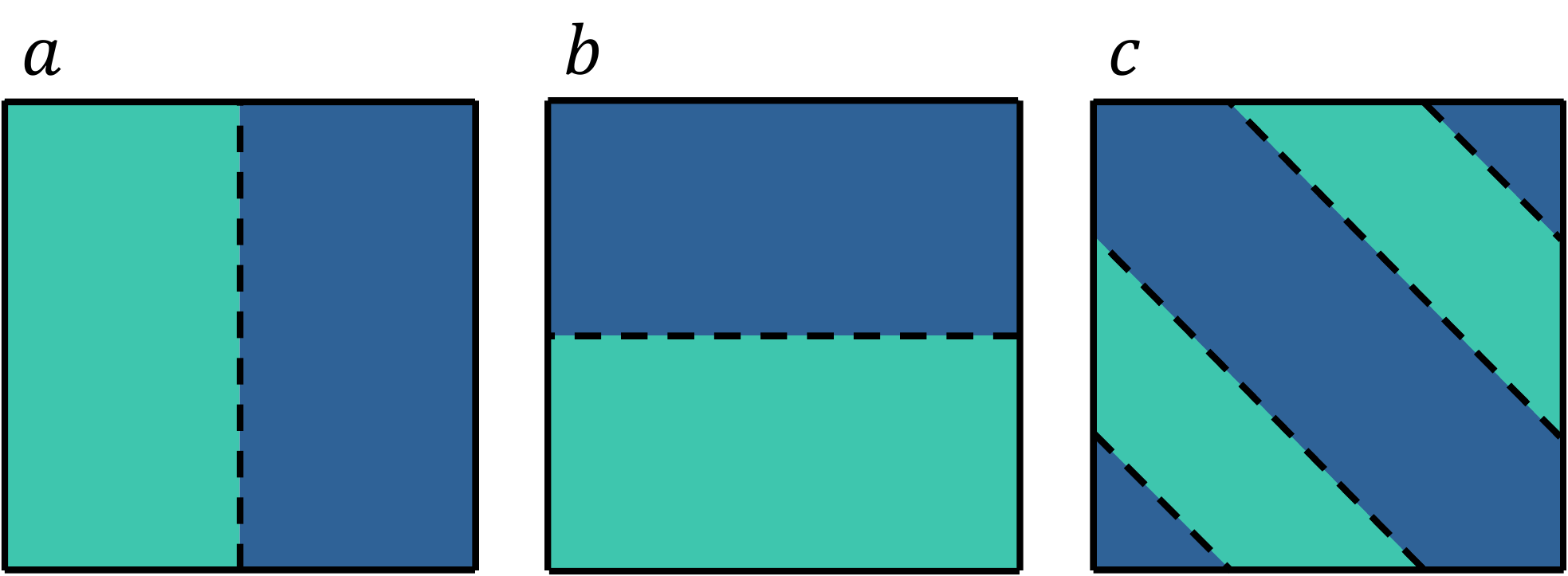}
    \caption{Partitions across which the second Renyi entropy is calculated.
    Subplots $a$, $b$, and $c$ correspond to the partitions used to obtain $S_{\leftrightarrow}$, $S_{\updownarrow}$, and $S_{\diagarrow}$, respectively.}
    \label{fig:entropy_partitions}
\end{figure}

To obtain the modular $S$ matrix, we follow the approach laid out in Ref.~\onlinecite{zhang2012quasiparticle}.
The first step is to use the penalty-function method \cite{wheeler2024ensemble} to find the number of degenerate ground states on the torus.
The ground-state degeneracy is the number of quasiparticle types.
For the $\nu=1/3$ state, there are three anyon types: the vacuum/identity ($e$), the quasihole ($qh$), and the quasielectron ($qe$). 
Therefore, we expect the ground state degeneracy to be three.
Call the (orthonormal) degenerate ground states we converge to $\ket{\xi_i}$, $i \in \{1, 2, 3\}$.
An arbitrary state in the degenerate subspace is a linear combination of the $\ket{\xi_i}$:
\begin{align}
    \ket{\psi} &= \sum_{i}\alpha_{i}\ket{\xi_{i}}.
\end{align}
Next, one finds the linear combination $|\Xi^{\leftrightarrow}_{1}\rangle$ that minimizes the second Renyi entropy, $S_{R}^{\leftrightarrow}= \ln[\text{Tr}(\rho_{A}^{2})]$, where $\rho_{A}$ is the reduced density matrix in a subregion $A$ of the torus:
\begin{align}
    |\Xi^{\leftrightarrow}_{1}\rangle=\text{argmin}_{|\psi\rangle} S_{R}^{\leftrightarrow}(\psi).
\end{align}
The symbol $\leftrightarrow$ refers to partitioning the torus horizontally, as shown in Fig.~\ref{fig:entropy_partitions}$a$.
The entropy is calculated using the method of Ref.~\onlinecite{tubman2012renyi}, which we summarize in the Supplementary Material.
One then repeats the entropy minimization process to find a three-state basis, enforcing orthogonality with previous states:
\begin{align}
    |\Xi^{\leftrightarrow}_{2}\rangle &= \text{argmin}_{|\psi\rangle|\langle\Xi^{\leftrightarrow}_{1}|\psi\rangle=0}S_{R}^{\leftrightarrow}(\psi),\\
    |\Xi^{\leftrightarrow}_{3}\rangle &= \text{argmin}_{|\psi\rangle|\langle\Xi^{\leftrightarrow}_{1}|\psi\rangle=\langle\Xi^{\leftrightarrow}_2|\psi\rangle=0}S_{R}^{\leftrightarrow}(\psi).
\end{align}
Similar bases are constructed by partitioning the system vertically, as shown in Fig.~\ref{fig:entropy_partitions}$b$, to obtain $|\Xi^{\updownarrow}_{i}\rangle$, and diagonally, as shown in Fig.~\ref{fig:entropy_partitions}$c$, to obtain $|\Xi^{\diagarrow}_{i}\rangle$. 
The modular $S$ matrix is then given by~\cite{zhang2015general}
\begin{align}
    S &= \{R[(\bar{U}^{(\updownarrow)})^{-1}\bar{U}^{(\leftrightarrow)}]\}^{-1}R[(\bar{U}^{(\updownarrow)})^{-1}\bar{U}^{(\diagarrow)}]\nonumber\\
    &\times R[(\bar{U}^{(\diagarrow)})^{-1}\bar{U}^{(\leftrightarrow)}],\label{eq:S}
\end{align}
where $R[X]$ denotes the matrix obtained after left and right multiplying $X$ by the diagonal matrices of complex phases required to make the first row and column of $X$ real and positive.
The $\bar{U}$ matrices are defined as 
\begin{align}
    |\Xi_{b}^{(\alpha)}\rangle &= \bar{U}^{(\alpha)}_{ab}|\xi_{a}\rangle.
\end{align}
Note that although $\bar{U}_{ab}^{(\alpha)}$ is not invariant under $U(1)$ transformations of $|\xi_{a}\rangle$, the resulting $S$ matrix is.


The $S$ matrix computed from our 8-electron simulation was:
\begin{widetext}
\begin{align}
  S = \frac{1}{\sqrt{3}}&\begin{pmatrix}
        1 & 1 & 1\\
        1 & e^{i 2\pi/3} & e^{-i 2\pi/3}\\
        1 & e^{-i 2\pi/3} & e^{i 2\pi/3}
    \end{pmatrix}
    +
    \begin{pmatrix}
        +0.006+0.001i & -0.032-0.002i & +0.017+0.000i\\
        +0.006-0.001i & +0.006+0.004i & -0.020+0.039i\\
        -0.020-0.001i & +0.016-0.025i & -0.005-0.017i
    \end{pmatrix}\nonumber\\
    \pm& 
    \begin{pmatrix}
        0.035 & 0.025 & 0.034\\
        0.032 & 0.043 & 0.034\\
        0.032 & 0.045 & 0.042
    \end{pmatrix}
    \pm
    \begin{pmatrix}
        0.002 & 0.003 & 0.002\\
        0.002 & 0.029 & 0.034\\
        0.003 & 0.036 & 0.028
    \end{pmatrix}i\label{eq:result_s}
\end{align}
\end{widetext}
where the first term is the expected result, the second term is the deviation from the expected result, and the last two terms are the statistical error of the mean calculated by computing the $S$ matrix 17 times.
Let the anyon ordering corresponding to the rows and columns of the $S$ matrix be $e$, $a$, $b$ (the entanglement interferometry method only gives the $S$ matrix up to a trivial permutation of the anyons' ordering sequence, so it is not possible to know whether $a$ is the $qh$ and $b$ the $qe$ or vice-versa).
The total quantum dimension is estimated to be $D = 1 / S_{ee}=\sqrt{3} + (0.001-0.004i) \pm 0.106\pm0.000i$.
The other quantum dimensions can be calculated using $d_{c} = DS_{ce}$ and come out to $d_{a}=1.016-0.004i\pm0.104\pm0.005i$, $d_{b}=0.971-0.004i\pm 0.101\pm0.008i$ (note that we have assumed $d_{e}=1$).
For a list of the computed fusion rules, see the Supplementary Material.
The phases obtained by braiding different anyon types around each other were calculated to be $\theta_{a,a}=2\pi/3 -0.014\pm0.079$, $\theta_{b,b}=2\pi/3+0.021\pm0.076$, and $\theta_{a,b}=-2\pi/3 -0.013\pm0.051$ (obtained from averaging $\arg(S_{a, b})$ and $\arg(S_{b, a})$).
As shown in the Supplementary Material, $S$ matrices computed for smaller systems also yield correct results, even for the tiny 4-electron system.
This is an important advantage of entanglement interferometry; it seems to produce the correct topological order even for very small systems, which would probably not be the case for entanglement scaling and entanglement spectroscopy.
We hope this will also be the case when studying more exotic phases, including non-Abelian systems.


In conclusion, this paper reported neural-network-based variational Monte Carlo simulations of the ground state of a FQHE system at filling fraction $\nu=1/3$ in the torus, confirming that small modifications of existing neural wavefunctions make good ansatze for the quantum Hall problem.
Using a penalty-function method \cite{wheeler2024ensemble}, we demonstrated numerically that the ground state is threefold degenerate, corresponding to the three anyon types in the theory: the vacuum/identity $e$, the quasihole $qh$ and the quasielectron $qe$.
The three ground states in the torus were then used to extract the modular $S$ matrix through entanglement interferometry~\cite{zhang2012quasiparticle, zhang2015general}.
The modular $S$ matrix contains topological information about the emergent anyonic excitations and thus identifies the topological field theory corresponding to the ground state.
We also obtained the quantum dimensions, fusion rules, and braid phases of the anyons.
This was all done in the continuum with Landau level mixing, completely ab initio and assumption-free.
We find that the topological anyon properties match those predicted by Wilczek~\cite{arovas1984fractional}.

More generally, we have extended an assumption-free approach for calculating anyonic quasiparticle properties in lattice models to the continuum.
Although we have investigated a system with known topological order, there is, in principle, no obstruction to using the same method for systems with unknown or debated topological order.
This work has the potential to shed light on the nature of fractional quantum Hall states at other filling fractions and higher Landau levels, and may help understand the properties of other systems in which interactions lead to emergent anyonic excitations.


\begin{acknowledgments}
We thank Frank Schindler for helpful discussions.
APF's PhD work is supported by the UK Engineering and Physical Sciences Research Council under grant EP/W524323/1.
This work was supported by a UKRI Future Leaders Fellowship MR/Y017331/1.
The Gauss Centre for Supercomputing e.V.\ (www.gauss-centre.eu) is acknowledged for providing computing time through the John von Neumann Institute for Computing (NIC) on the GCS Supercomputer JUWELS at J\"{u}lich Supercomputing Centre (JSC), and the EUROfusion consortium for providing computing time on the Leonardo Supercomputer at CINECA in Bologna.
The EUROfusion Consortium is funded by the European Union via the Euratom Research and Training Programme
(Grant Agreement No 101052200 — EUROfusion).
Views and opinions expressed are however those of the authors only and do not necessarily reflect those of the European Union or the European Commission.
Neither the European Union nor the European Commission can be held responsible for them. 
The authors acknowledge the use of resources provided by the
Isambard-AI National AI Research Resource (AIRR)~\cite{mcintosh2024isambard}.
Isambard-AI is operated by the University of Bristol and is funded by the UK Government’s Department for Science, Innovation and Technology (DSIT) via UK Research and Innovation; and the Science and Technology Facilities Council [ST/AIRR/I-A-I/1023].
T.N. acknowledges support from the Swiss National Science Foundation through a Quantum grant (20QU-1\_225225).
\end{acknowledgments}


\bibliography{ref}



\section{Supplementary Material}
\label{sec:supplementary}

\section{Hyperparameters}
The network and training hyperparameters are shown in table \ref{table:hyperparameters}
\begin{table}[h!]
\centering
\begin{tabular}{c c}
  \hline
  Parameter & value\\
  \hline
  Pretraining iterations & 0\\
  Learning rate & $0.15\times (1 + t/10000)^{-0.5}$\\
  Determinants & 32\\
  Dimensions & 2\\
  Clip from median & True\\
  Complex output & True\\
  Jastrow hidden layers & 3\\
  Jastrow hidden width & 64\\
  Jastrow activations & tanh\\
  Landau level cutoff $N_{\text{LL}}$ & 3\\
  \hline
\end{tabular}
\caption{Table of hyperparameters used. Any hyperparameter not referenced was taken to be the default value~\cite{ferminet_github}. \label{table:hyperparameters}
}
\end{table}\\[3.0em]

\section{Fusion Rules}
\label{chap:fusion}
Given the modular $S$ matrix, the fusion rules can be calculated using the Verlinde formula,
\begin{align}
    a\times b &= \sum_{c}N_{ab}^{c}c,\\
    N_{ab}^{c} &= \sum_{d}\frac{S_{ad}S_{bd}S_{dc}^{*}}{S_{1d}},
\end{align}
where $a$, $b$, $c$, $d$ are anyon types, and $\times$ denotes fusion.
If $a\times b=c$, then anyons $a$ and $b$ can fuse into anyon $c$.
If $a\times b=c+d$, then anyons $a$ and $b$ can fuse into either anyon $c$ or anyon $d$.
If an anyon has multiple fusion channels, then it must be non-Abelian.
From the $N=8$ PS+J calculation, the fusion rules for the $\nu=1/3$ state are found to be (large numbers in bold)
\begin{widetext}
\begin{align}
    e\times e&= (\bm{1.000} + 0.000i)e + (0.000 + 0.000i)a + (0.000 + 0.000i)b,\\
    e\times a &= (0.000 - 0.000i)e + (\bm{1.000} + 0.000i)a + (0.000 + 0.000i)b,\\
    e\times b &= (0.000 + 0.000i)e + (0.000 + 0.000i)a + (\bm{1.000} + 0.000i)b,\\
    a\times a &= (0.056 - 0.003i)e + (0.005 + 0.065i)a + (\bm{0.997} - 0.070i)b,\\
    a\times b &= (\bm{0.974} - 0.003i)e + (0.009 - 0.072i)a + (-0.040 + 0.073i)b,\\
    b\times b &= (-0.031 - 0.006i)e + (\bm{0.992} + 0.076i)a + (-0.036 - 0.081i)b.
\end{align}
\end{widetext}
In other words, two quasiholes fuse into a quasielectron (plus a hole, which is a local particle), and two quasielectrons fuse into a quasihole (plus an electron).
The fusion of any anyon $c$ with the vacuum is $c$ as expected.
The average statistical error of the mean of the elements of the fusion $N$ tensor is 0.049.

\section{Evaluating the Renyi Entropy}
\label{chap:entropy_calc}
The method employed we use to compute the quasiarticle properties is based on calculating and minimizing the Renyi entropy.
The $n$th Renyi entropy for a system split into 2 partitions $A$ and $B$ is defined as 
\begin{align}
    (S_{n})_{R} &= \frac{1}{n-1}\ln[\text{Tr}(\rho_{A}^{n})],
\end{align}
where $\rho_{A}=\text{Tr}_{B}(|\psi\rangle\langle\psi|)$ is the reduced density matrix in region $A$.
Note that the von Neumann entropy is recovered in the limit $n\to 1$.
In the main text, we use the notation $S_{R}=(S_{2})_{R}$.
The first paper to show how to estimate this quantity in quantum Monte Carlo simulations of lattice systems was Ref.\ \cite{hastings2010measuring}; the technique was extended to continuum systems in Ref.\ \cite{tubman2012renyi}.
The required estimator takes the form:
\begin{align}
    e^{-(S_{2})_{R}} &= \left\langle\frac{\psi(\text{SWAP}_{A}(\bm{R}_{1}, \bm{R}_{2}))\psi(\text{SWAP}_{A}(\bm{R}_{2}, \bm{R}_{1}))}{\psi(\bm{R}_{1})\psi(\bm{R}_{2})}\right\rangle,
\end{align}
where $\bm{R}_{1},\bm{R}_{2}\sim|\psi|^{2}$.
The notation means the following.
Take two random samples, $\bm{R}_1$ and $\bm{R}_2$, from $|\psi|^{2}$.
The $\text{SWAP}_{A}(\bm{R}_{1}, \bm{R}_{2})$ operator takes all of the electron positions in $\bm{R}_{1}$ and exchanges those within region $A$ with the electron positions in region $A$ found in $\bm{R}_{2}$.
If $\bm{R}_{1}$ and $\bm{R}_{2}$ do not have the same number of electrons in region $A$, the contribution of that pair of walkers to the expectation value is zero. 
The entropy was minimized with the ADAM optimizer, and the gradients were calculated with automatic differentiation in JAX.

\section{$S$ matrix for smaller systems}
The modular $S$ matrices for all of the Psiformer + Jastrow (PS+J) runs mentioned in the main text can be found below. 
We define $\Delta S$ as the difference between our numerical result and the expected result as given by the first term in Eq.\ \ref{eq:result_s}.
Note that while the error for $N=8$ was obtained as the average of 16 calculations of the $S$ matrix, only 1 was used for $N=4$, 3 were used for $N=5$ and 4 for $N=6$, so their deviations from the expected result (and their uncertainties) are larger.
\begin{widetext}
\begin{align}
    \Delta S_{N=4} &= \begin{pmatrix}
        -0.006+0.000i & -0.005-0.000i & +0.011-0.000i\\
        -0.025+0.001i & +0.024+0.030i & +0.009-0.016i\\
        +0.030-0.001i & -0.011+0.018i & -0.010-0.031i
    \end{pmatrix}\nonumber\\
    \Delta S_{N=5} &= \begin{pmatrix}
        -0.028-0.001i & +0.021-0.002i & +0.004+0.002i\\
        +0.025-0.002i & -0.013-0.020i & +0.030+0.007i\\
        +0.001+0.002i & +0.034+0.001i & +0.007+0.012i\\
    \end{pmatrix}\nonumber\\
    \Delta S_{N=6} &= \begin{pmatrix}
        -0.014-0.001i & -0.020+0.001i & +0.031-0.000i\\
        +0.001-0.001i & -0.006+0.022i & +0.021+0.022i\\
        +0.011+0.002i & +0.050-0.015i & -0.033-0.031i
    \end{pmatrix}\nonumber\\
    \Delta S_{N=8} &= \begin{pmatrix}
        +0.006+0.001i & -0.032-0.002i & +0.017+0.000i\\
        +0.006-0.001i & +0.006+0.004i & -0.020+0.039i\\
        -0.020-0.001i & +0.016-0.025i & -0.005-0.017i
    \end{pmatrix}\nonumber
\end{align}
\end{widetext}

\section{Convergence}
\label{chap:convergence}
\begin{figure}[h]
    \centering
    \includegraphics[scale = 0.5]{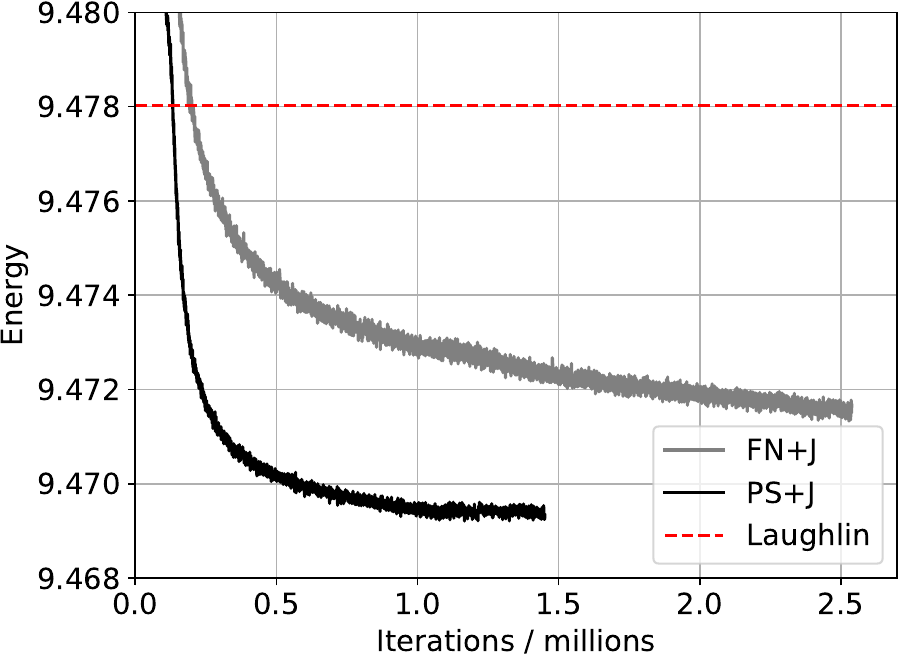}
    \caption{Learning curve for the loss function against the number of iterations of a 6-electron calculation. 
    The dashed line represents the energy of the Laughlin wavefunction in the torus.}
    \label{fig:learning_curve}
\end{figure}

Unfortunately, and for reasons not yet clear to the authors, the number of iterations needed to converge a fractional quantum Hall calculation is much greater than that of chemical systems of the same number of electrons.
An example of the learning curve for 6 electrons is shown in figure \ref{fig:learning_curve}.
The authors believe the slow convergence is due to the implementation of the boundary conditions through the envelope; this is something that will be investigated further.


\end{document}